%% file: spCV.tex
\documentclass[preprint,12pt]{elsarticle}
\usepackage[OT1]{fontenc}
\usepackage{amsthm,amsmath,amsfonts,amssymb}%
\usepackage[colorlinks,citecolor=blue,urlcolor=blue]{hyperref}%

\usepackage{latexsym}%
\usepackage{mathtools}
\usepackage{mathrsfs}
\usepackage{dsfont}
\usepackage{graphicx}%
\usepackage{caption}
\usepackage{subcaption}
\usepackage{adjustbox}
\usepackage{lscape} %Querformat
\usepackage[all]{xy}	% used to create trees, graphs, (commutative) diagrams, and similar things

\usepackage{array}%
\usepackage{tabularx}
\newcolumntype{L}[1]{>{\raggedright\arraybackslash}p{#1}} % linksbündig mit Breitenangabe
\newcolumntype{C}[1]{>{\centering\arraybackslash}p{#1}} % zentriert mit Breitenangabe
\newcolumntype{R}[1]{>{\raggedleft\arraybackslash}p{#1}} % rechtsbündig mit Breitenangabe

\usepackage{bm}	%Allows use of bold greek letters in math mode

\usepackage{nicefrac} %package for nice fractions
\usepackage{booktabs}	%setzen von qualitativ hochwertigen horizontalen Strichen (engl. rules) in Tabellen
\def\bSig\bm{\Sigma}

\newcommand{\beqo}{\[}
\newcommand{\eeqo}{\]}
\newcommand{\beqq}{\begin{equation}}
\newcommand{\eeqq}{\end{equation}}

\let\oldsqrt\sqrt
\def\sqrt{\mathpalette\DHLhksqrt}
\def\DHLhksqrt#1#2{%
\setbox0=\hbox{$#1\oldsqrt{#2\,}$}\dimen0=\ht0
\advance\dimen0-0.2\ht0
\setbox2=\hbox{\vrule height\ht0 depth -\dimen0}%
{\box0\lower0.4pt\box2}}

\newcommand{\R}{\mathbb{R}}

\newcommand{\calJ}{\mathcal{J}}

\newcommand{\calD}{\mathcal{D}}

\newcommand{\calL}{\mathcal{L}}

\newcommand{\calR}{\mathcal{R}}
\newcommand{\simiid}{\stackrel{\text{i.i.d.}}{\sim}}
\newcommand{\veps}{{\varepsilon}}

\newcommand{\te}{{\theta}}

\newcommand{\btheta}{\bm \theta}

\newcommand{\bnu}{\bm \nu}

\newcommand{\bbeta}{\bm \beta}

\newcommand{\hattheta}{\widehat{\bm \theta}}
\newcommand{\hattau}{\widehat{\bm \tau}}
\newcommand{\hatnu}{\widehat{\bm \nu}}

\newcommand{\bU}{{\bm{U}}}
\newcommand{\bu}{{\bm{u}}}

\newcommand{\bx}{{\bm{x}}}
\newcommand{\by}{{\bm{y}}}
\newcommand{\bY}{{\bm{Y}}}

\newcommand{\Fzi}{g_z^{-1}}
\newcommand{\itau}{g^{-1}_{\tau}}
\newcommand{\Rsq}{\text{R}^2}
\newcommand{\Rsqa}{\text{R}^2_\text{adj}}
\newcommand{\vN}[2]{\mathcal{N}\left(#1,#2\right)}

\newcommand{\vU}[2]{\mathcal{U}\left(#1,#2\right)}

\newcommand{\ov}{\overline}
\newcommand{\wh}{\widehat}

\newcommand{\sepl}{\,|\,}
\newcommand{\sepi}{;}

\newcommand{\half}{\frac{1}{2}}

\newcommand{\iN}[2]{#1=1,\ldots,#2}
\newcommand{\xn}[2]{#1_1,\ldots,#1_{#2}}

\newcommand{\sumtN}{\sum_{t=1}^N}
\newcommand{\sumsd}{\sum_{s=1}^d}
\newcommand{\prodtN}{\prod_{t=1}^N}
\newcommand{\prodsd}{\prod_{s=1}^d}

\newcommand{\T}{^\top}	%Transponieren

\newcommand{\lnf}[1]{\ln\left(#1\right)}

\newcommand{\edge}{\mathcal{E}}	%Edges
\newcommand{\tree}{\mathcal{T}}	%Tree

\newcommand{\vine}{\mathscr{V}}	%Vine
\newcommand{\cvc}{\mbox{CVC}}	%C-vine collection
\newcommand{\Cop}[2]{C_{#1}\left(#2\right)}	%Copula
\newcommand{\cop}[2]{c_{#1}\left(#2\right)}	%Copula density
\newcommand{\Lik}[3]{\calL_{#1}\left(#2\sepi#3\right)}	%likelihood
\renewcommand{\ll}[3]{\ell_{#1}\left(#2\sepi#3\right)}	%log-likelihood
\newcommand{\cll}[3]{c\ell_{#1}\left(#2\sepi#3\right)}	%composite log-likelihood
\newcommand{\partialderiv}[1]{\frac{\partial}{\partial #1}}

\begin{document}

\begin{frontmatter}

\title{Spatial composite likelihood inference using local C-vines}%\tnoteref{t1}}
\author[tum]{Tobias Michael~Erhardt\corref{cor1}\fnref{fn1}}
\ead{tobias.erhardt@tum.de}
\author[tum]{Claudia~Czado}
\ead{cczado@ma.tum.de}
\author[tum]{Ulf~Schepsmeier}
\ead{ulf.schepsmeier@tum.de}

\cortext[cor1]{Corresponding author, Phone: +49 89 289 17425}
%\cortext[cor2]{Principal corresponding author}
\fntext[fn1]{Supported by the TUM Graduate School's Graduate Center International Graduate School of Science and Engineering (IGSSE).}
\address[tum]{Zentrum Mathematik, Technische Universit\"at M\"unchen, Boltzmannstr. 3, 85748 Garching, Germany}

%  put the summary for your paper here

\begin{abstract}

We present a vine copula based composite likelihood approach to model spatial dependencies, which allows to perform prediction at arbitrary locations. This approach combines established methods to model (spatial) dependencies. On the one hand the geostatistical concept utilizing spatial differences between the variable locations to model the extend of spatial dependencies is applied. On the other hand the flexible class of C-vine copulas is utilized to model the spatial dependency structure locally. These local C-vine copulas are parametrized jointly, exploiting an existing relationship between the copula parameters and the respective spatial distances and elevation differences, and are combined in a composite likelihood approach. The new methodology called spatial local C-vine composite likelihood (S-LCVCL) method benefits from the fact that it is able to capture non-Gaussian dependency structures. The development and validation of the new methodology is illustrated using a data set of daily mean temperatures observed at $73$ observation stations spread over Germany. For validation continuous ranked probability scores are utilized. Comparison with two other approaches of spatial dependency modeling introduced in yet unpublished work of Erhardt, Czado and Schepsmeier (2014) (see \cite{erhardt14}) shows a preference for the local C-vine composite likelihood approach.

\footnotetext[0]{C-vine = canonical vine, R-vine = regular vine, (S-)LCVCL = (spatial) local C-vine composite likelihood, SC = spatial local C-vine composite likelihood approach, SV = spatial R-vine model, SG = spatial Gaussian model}
\end{abstract}


%\maketitle
\end{frontmatter}

%\clearpage

%%
%% Start line numbering here if you want
%%
%\linenumbers

% Einbinden der einzelnen Kapitel
\input{Sections/Introduction.tex}
\input{Sections/Vines.tex}

\input{Sections/Data.tex}

\input{Sections/Composite.tex}

\input{Sections/Validation.tex}

\input{Sections/Conclusions.tex}

%\appendix
%
%\section{Appendix section}\label{app}
%
%\subsection{Appendix subsection}

\section*{Acknowledgments}

%The authors gratefully acknowledge the helpful comments of the referees, who further improved the manuscript.
The numerical computations were performed on a Linux cluster supported by DFG grant INST 95/919-1 FUGG. \vspace*{-8pt}

%\section*{Supplementary Materials}
%
%Appendix A, referenced in Section \ref{sec:app:SVpred} and Section \ref{sec:classical} and the figures referenced in Sections \ref{sec:rvines}, \ref{sec:app:SVfit}, \ref{sec:SVpred:results} and \ref{sec:classical} are available with this paper.\vspace*{-8pt}

\section*{References}

\bibliographystyle{elsarticle-num}
\bibliography{literature}

\end{document}

%% file: Sections/Introduction.tex
\section[Introduction]{Introduction}\label{sec:intro}  % neues Kapitel mit Namen "Introduction"

%Aufbau:
%Motivation -> Problem
%Kurz: Was machst du? -> Ziel der Arbeit
%Literatur -> bisherige Ansätze
%etwas ausführlicher: Wie wirst du vorgehen?

Research on spatial model building has a long history and found application in a broad field of disciplines like for instance in climatology, ecology, epidemiology, forestry, geology and hydrology. This paper presents a new vine copula based approach in the area of spatial dependency modeling, which is combining local vine models. It utilizes available spatial information to set up a parametric composite likelihood, the spatial local C-vine composite likelihood (S-LCVCL). Parameter estimation and prediction methods for prediction at arbitrary locations are developed. The presented methodology is applied to model the spatial dependencies of climatic mean temperature time series.
%The required statistical inference methods are developed and a prediction method which allows prediction at arbitrary locations is presented.

The conventional distributions used in the context of spatial dependency modeling are multivariate Gaussian distributions. For example \cite{benth07} model daily temperature averages using a Gaussian random field based spatial-temporal model. Also \cite{hu13} assume a Gaussian dependence structure. They use systems of stochastic partial differential equations to spatially model the dependence of temperature and humidity. However Gaussian distributions are not appropriate for the modeling of arbitrary multivariate data, since they are subject to shortcomings regarding their flexibility in terms of symmetry and extreme dependence. For that reason we are going to apply vine copulas, which are not prone to these limitations.

Copulas are distribution functions $C:[0,1]^d\to[0,1]$. They are the link between a multivariate distribution function $F$ and the respective marginals ($\xn{F}{d}$) and incorporate all dependency information (see \cite{sklar59}). In particular for any distribution of a continuous random vector $\bY=(Y^1,\ldots,Y^d)\T\in\R^d$ we have $F(\by)=C\left(F_1(y^1),\ldots,F_d(y^d)\right)$.

Vine copulas are flexible $d$-dimensional copulas composed out of bivariate copulas, which are well understood and easy to estimate (see \cite{aas09, stoeber12}). An introduction of a subclass called C-vine copulas will be given in Section \ref{sec:cvines}.

Different vine copula based spatio-temporal models are provided in \cite{erhardt14} and \cite{graeler14}. Both approaches employ the geostatistical idea to exploit the dependency information contained in the spatial distances between variable locations. The approach of \cite{graeler14} combines the geostatistical idea of binning with C-vine copulas to model spatio-temporal dependencies. In contrast our new approach and the spatial R-vine model presented in \cite{erhardt14} additionally allow to exploit supplementary spatial covariates like for instance elevation differences, to model the spatial dependencies implied by the vine specification.

The new modeling approach will be illustrated using daily mean temperature time series collected over the period 01/01/2010-12/31/2012 by the German Meteorological Service (Deutscher Wetterdienst).

The innovation presented is a vine copula based approach to model spatial dependencies locally and combine these local models in a composite likelihood. The resulting spatial local C-vine composite likelihood (S-LCVCL) will be introduced in Section \ref{sec:composite}. It relies on a joint parametrization of several local C-vines, which exploits the link of the model parameters to the available spatial information. Different distance and elevation difference based model specifications were taken into consideration. We focus on the most promising one among them. Maximum composite likelihood estimation of the involved parameters is presented and a method for prediction at an arbitrary location is illustrated.

For model validation and comparison in Section \ref{sec:appl} we additionally consider two spatial dependency models, the spatial R-vine model and the spatial Gaussian model introduced in \cite{erhardt14}. Adequate scores, calculated based on a validation data set consisting of time series for $19$ additional locations, allow to compare different predictions. Moreover we see that the presented spatial local C-vine composite likelihood method yields a distinct reduction in computation time, compared to the two other approaches.

Further applications in different areas and modifications of the presented methodology are feasible. For instance the modeling of biomass or pollutants using a spatial local C-vine composite likelihood approach and variants thereof might lead to new insights.

%% file: Sections/Vines.tex
\section[Canonical vine copulas]{Canonical vine copulas}\label{sec:cvines}

This section gives a short introduction to vine copulas arising from \emph{canonical vines (C-vines)}, which will be the components of our proposed spatial local C-vine composite likelihood.

Vine copulas trace back to ideas of Joe \cite{joe96}. They are multivariate copulas arising from \emph{pair-copula constructions} (see \cite{aas09}). Pair-copula constructions are decompositions of $d$-dimensional (copula) densities into a product of $\nicefrac{d(d-1)}{2}$ bivariate copula densities. These decompositions enable easy estimation of these highly flexible vine copulas. There exists a pair-copula construction for each multivariate copula density, however it does not has to be unique in general. A tool which helps to organize pair-copula constructions is the regular vine (R-vine) introduced in \cite{bedfordcooke01,bedfordcooke02}. We focus on a special case of these R-vines, so called canonical vines (C-vines).

\begin{figure}[htb]
	\centerline{
		\entrymodifiers={++[o][F]}
		%\SelectTips{cm}{}
		\xymatrixrowsep{0pc}\xymatrix @+2pc {
		*{\tree_1}					& *\txt{}							&	*{\tree_2}	& *\txt{}									&	*{\tree_3}									\\
		2	\ar@{-}[d]_{1,2}	& 3 \ar@{-}[dl]_{1,3}	&	1,2					& 1,3 \ar@{-}[l]_{2,3;1}	& 2,3;1 \ar@{-}[d]^{3,4;1,2}	\\
		1										& 4 \ar@{-}[l]_{1,4}	&	*\txt{}			& 1,4 \ar@{-}[ul]^{2,4;1}	& 2,4;1												\\
		}
	}
	\caption{Four dimensional C-vine.}
	\label{fig:cvine}
\end{figure}
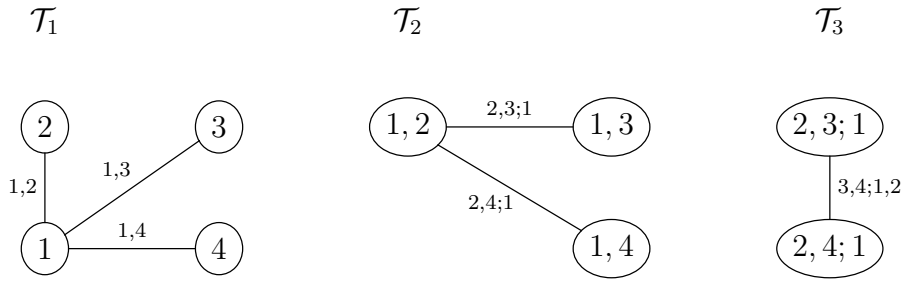

Vines in general are nested sets of trees $\vine=(\xn{\tree}{d-1})$, which capture the structure of pair-copula constructions. We now explain the concept of a C-vine in a general $d$-dimensional setting (see \cite{czado12}). To ease the understanding we illustrate a four dimensional C-vine in Figure \ref{fig:cvine}. Let us now w.l.o.g. consider variables labeled from $1$ to $d$. The first tree $\tree_1$ of a C-vine represents the dependence of the first root node variable labeled $1$ on all other involved variables labeled $2,\ldots,d$, which are depicted as further tree nodes. These dependencies are indicated by the tree edges between the respective nodes, which are labeled $\left\{1,2\right\},\ldots,\left\{1,d\right\}$. The nodes of the trees $\tree_l$, $l=2,\ldots,d-1$, coincide with the edges of the respective previous tree $\tree_{l-1}$. The variable corresponding to $l$ is chosen to be the root variable of the tree $\tree_l$. The edges between the root node $\left\{l-1,l;1,\ldots,l-2\right\}$ and the remaining tree nodes are conditioned on the root variables $1,\ldots,l-1$ of the previous trees $\tree_1,\ldots,\tree_{l-1}$, i.e. they are labeled $\left\{l,l+1;1,\ldots,l-1\right\},\ldots,\left\{l,d;1,\ldots,l-1\right\}$ and represent the conditional dependence of the $l$th root variable $l$ and the variables $l+1,\ldots,d$ given the variables $1,\ldots,l-1$.

The structure of a C-vine is uniquely determined by the ordering of the root variables $\calR\coloneqq\left\{1,\ldots,d\right\}$. The construction described above induces a star structure for every tree $\tree_1,\ldots,\tree_{d-1}$. Such a structure appears to be suitable in every setting where a natural ordering of the variables under consideration is given. An example would be a spatial arrangement of variables. One may select a variable of interest and order the remaining variables according to their spatial distance with the selected variable.

%So called regular vines (R-vines) are used to organize pair-copula constructions. An R-vine $\vine$ is a nested set of trees $\vine=(\xn{\tree}{d-1})$, for which certain conditions hold (see \cite{bedfordcooke01}):
%\begin{enumerate}
	%\item $\tree_1=(\vrtx_1,\edge_1)$ is a tree with vertices $\vrtx_1=\{1,\ldots,d\}$ and edges $\edge_1$.
	%\item $\tree_l=(\vrtx_l,\edge_l)$ is a tree with vertices $\vrtx_l=\edge_{l-1}$ and edges $\edge_l$, for all $l=2,\ldots,d-1$.
	%\item For all vertex pairs in $\vrtx_l$ connected by an edge $e\in\edge_l$, $l=2,\ldots,d-1$, the corresponding edges in $\edge_{l-1}$ have to share a common vertex (\emph{proximity condition}).
%\end{enumerate}
%
%For the denotation of vine edges we follow \cite{czado10}. We denote an edge $e \in \edge_l$, $\iN{l}{d-1}$ by $i(e),j(e);\calD_e$, where $i(e)$ and $j(e)$ constitute the \emph{conditioned set} $\calC_e=\{i(e), j(e)\}$ and $\calD_e$ is called \emph{conditioning set}.

Now we are interested in the pair-copula construction encoded by the C-vine tree sequence $\vine=(\xn{\tree}{d-1})$ corresponding to a multivariate copula density $c$ of some random vector $\bU=(U^1,\ldots,U^d) \in [0,1]^d$ with random components $U^1,\ldots,U^d \sim \vU01$. Each C-vine edge represents one of the bivariate building blocks of the pair-copula construction, i.e. a (parametric) bivariate copula density $c_{l,k;1:l-1}$ respectively a copula distribution function $C_{l,k;1:l-1}$ is assigned to each edge $\{l,k;1,\ldots,l-1\}$, where $k=l+1,\ldots,d$ and $l=1,\ldots,d-1$. The respective pair-copula construction is given by
\beqq\label{eq:cvdensity}
		c(\bu)=\prod_{l=1}^{d-1} \prod_{k=l+1}^{d} \cop{l,k;1:l-1}{C_{l|1:l-1}(u^{l}\sepl \bu^{1:l-1}),C_{k|1:l-1}(u^{k}\sepl \bu^{1:l-1})},
\eeqq
where $\bu^{\calJ} \coloneqq \left\{u^j: j\in\calJ\right\}$. The transformed variables $C_{j|1:l-1}(u^{j}\sepl \bu^{1:l-1})$, $j=l,\ldots,d$, $l=1,\ldots,d-1$, are calculated recursively according to \cite{joe96}, using the formula
%\beqq\label{eq:joe}
	%C_{j|1:l-1}(u^j \sepl \bu^{1:l-1}) =	\frac{\partial\ \Cop{l-1,j;\calJ_l}{C_{j|\calJ_l}(u^j \sepl \bu^{\calJ_l}), C_{l-1|\calJ_l}(u^{l-1} \sepl \bu^{\calJ_l})}}{\partial C_{l-1|\calJ_l}(u^{l-1} \sepl \bu^{\calJ_l})},
%\eeqq
%where $j=l,\ldots,d$ and $\calJ_l=1:l-2$.
\beqq\label{eq:joe}
	C_{j|1:l-1} =	\frac{\partial\ \Cop{l-1,j;1:l-2}{C_{j|1:l-2}, C_{l-1|1:l-2}}}{\partial C_{l-1|1:l-2}}, \quad j=l,\ldots,d,
\eeqq
where the arguments are omitted. The distributions $C_{j|1:l-1}$, $j=l,\ldots,d$, are conditional distributions obtained from the bivariate copula distribution $C_{l-1,j;\calJ_l}$. We implicitly made a \emph{simplifying assumption}, that the copula densities $c_{l,k;1:l-1}$ given in \eqref{eq:cvdensity} do not depend on the conditioning value $\bu^{1:l-1}$ other than through its arguments.

The number of parameters needed to parametrize each copula $C_{l,k;1:l-1}$, depends on the respective copula family denoted by $b_{l,k;1:l-1}$. For an overview of frequently used bivariate copula families we refer to \cite{cdvine13}.

%% file: Sections/Data.tex
\section{Mean Temperature Data}\label{sec:data}

We consider daily (01/01/2010-12/31/2012) mean temperature data in $^{\circ}\mathrm{C}$ gathered by the German Meteorological Service (Deutscher Wetterdienst) at $73$ observation stations scattered over Germany and splitted the data set into a training ($\iN{s}{54}$) and a validation data set ($s=55,\ldots,73$).
Hence our models are built on $d=54$ times $N=1096$ observations $y_t^s$ ($\iN{t}{N}$, $\iN{s}{d}$) of daily mean temperatures. The observations $\xn{y^s}{N}$ collected at the observation stations $\iN{s}{d}$ are considered to be realizations of time series $\xn{Y^s}{N}$, $\iN{s}{d}$.

\begin{figure}[htb]
	\centering
		\includegraphics[width=0.48\textwidth]{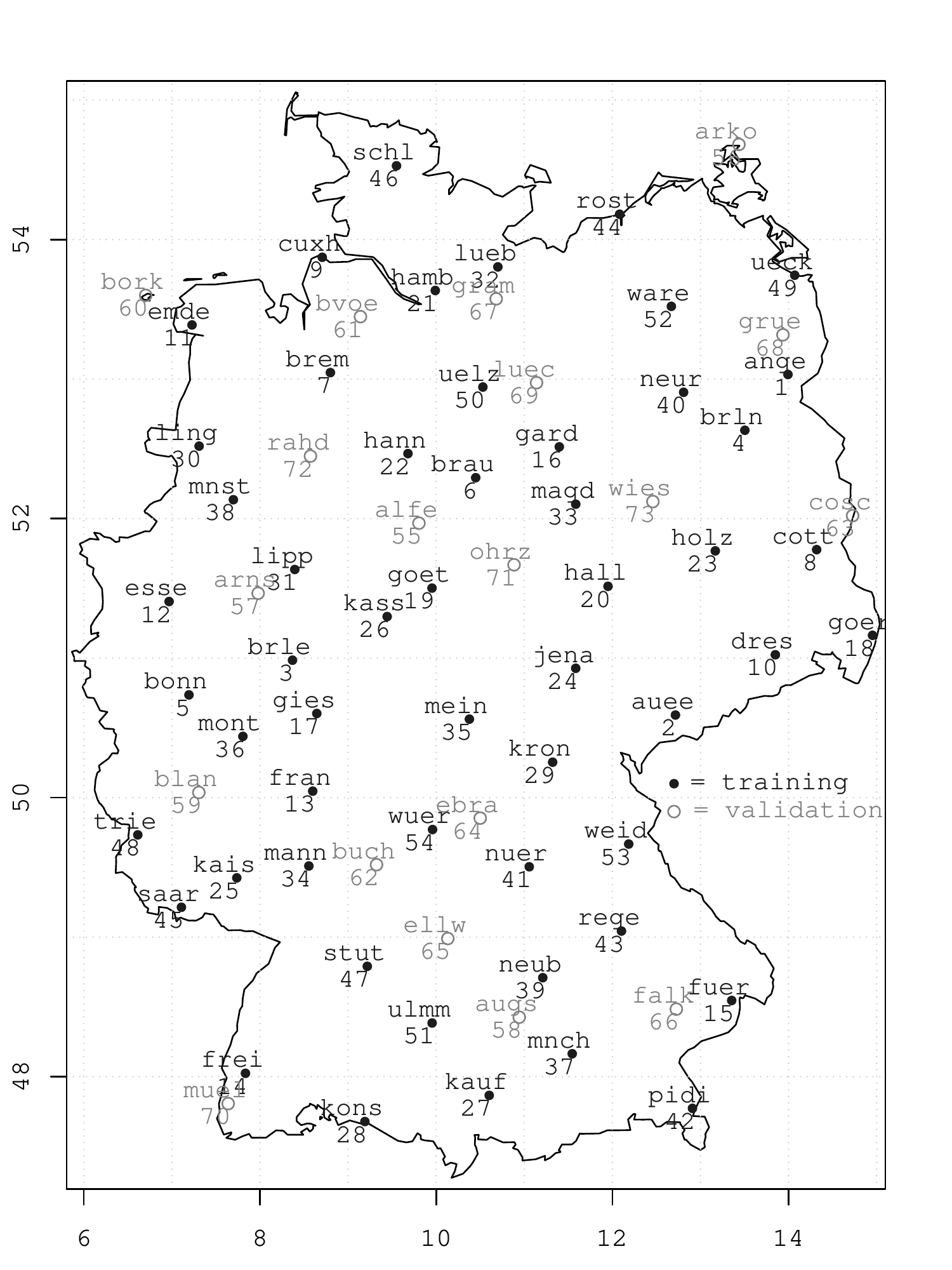}
	\caption[Observation stations]{Observation stations: Location of the $73$ observation stations in Germany (training ($1,\ldots,54$) and validation data ($55,\ldots,73$) with short name and ID).}
	\label{fig:stations}
\end{figure}

The locations (longitude, latitude and elevation) of all $73$ observation stations and the respective station names are provided in Table 3.1 and 5.9 of \cite{erhardt13} and are visualized in Figure \ref{fig:stations}.

%\begin{figure}[htb]
	%\centering
		%\includegraphics[width=0.495\textwidth]{Figures/ObsStations3.pdf}
		%\includegraphics[width=0.495\textwidth]{Figures/GermanySCVM.pdf}
	%\caption[Location of the $73$ observation stations in Germany and visualization of a (spatial) composite vine model.]{Left panel: Location of the $73$ observation stations in Germany (training ($1,\ldots,54$) and validation data ($55,\ldots,73$) with short name and ID). Right panel: Visualization of the structure of a (spatial) local C-vine composite likelihood for the training data set (for details see Section \ref{sec:CVM:app}).}
	%\label{fig:stations}
%\end{figure}

Since the raw data is not restricted to the unit hypercube $[0,1]^d$ with uniformly distributed margins, we have to transform it to so called \emph{copula data} $u_t^s\sim\vU01$, $\iN{s}{d}$, $\iN{t}{N}$, if we want to work with vine copula models. In a first step of the transformation we utilize a regression model $Y_t^s=g(t, \bx^s; \bbeta) + \veps_t^s$, $\veps_t^s \sim F^s$, where $\bx^s$ are some spatial covariates. The regression model should be able to adjust for spatial trends, seasonality effects and time dependencies in the margins. If the adjustment is appropriate, the respective residuals $\wh\veps_t^s \coloneqq y_t^s - g(t, \bx^s; \wh\bbeta)$, $\iN{t}{N}$, can be considered to be independent for each location $\iN{s}{d}$. In a second step they are transformed using the parametric marginal distributions $F^s$, calculating $u_t^s \coloneqq F^s(\wh\veps_t^s)$. Such transformations are called probability integral transform. We favor to apply such parametric probability integral transformations (see \cite{joe96b}) instead of empirical rank transformations (see \cite{genest95}), since we aim to predict on the original scale using the proposed marginal models.

%% file: Sections/Composite.tex
\section[Spatial dependency modeling using local C-vines]{Spatial dependency modeling using local C-vines}\label{sec:composite}

In \cite{erhardt14} we presented a straightforward vine copula based approach to model spatial dependencies by embedding spatial information into the parametrization of an R-vine model. In contrast we now utilize composite likelihood \cite{varin11} by joining local dependency models. The modeling of these local dependencies is carried out by C-vine copulas. The computations in this section and the forthcoming sections are based on the statistical software R \cite{r14} and the R-package \texttt{VineCopula} of Schepsmeier et al. \cite{VineCopula14}, providing substantial functionality for vine copulas.

\subsection{Local C-vine composite likelihood (LCVCL)}\label{sec:CVM}

%In a first step we are going to propose a local C-vine based composite likelihood approach for the purpose of spatial dependency modeling.
Since the local dependencies are modeled with C-vines, we require copula data $\bu^1,\ldots,\bu^d$ with $\bu^s=(u^s_{1},\ldots,u^s_{N})\T$, $\iN{s}{d}$. Hence our raw data $y_t^s$, $\iN{t}{N}$, $\iN{s}{d}$, has to be transformed appropriately. For this we utilize the marginal transformation developed in Section 3 of \cite{erhardt14}.

\begin{figure}[htb]
	\centerline{
		\entrymodifiers={++[o][F]}
		%\SelectTips{cm}{}
		\xymatrixrowsep{0pc}\xymatrix @+2pc {
		*{\tree_1^s}						& *\txt{}									&	*{\tree_2^s}	& *\txt{}												&	*{\tree_3^s}							\\
		p_s	\ar@{-}[d]_{s,p_s}	& q_s \ar@{-}[dl]_{s,q_s}	&	s,p_s					& s,q_s \ar@{-}[l]_{p_s,q_s;s}	& p_s,q_s;s \ar@{-}[d]^{q_s,r_s;s,p_s}	\\
		s												& r_s \ar@{-}[l]_{s,r_s}	&	*\txt{}				& s,r_s \ar@{-}[ul]^{p_s,r_s;s}	& p_s,r_s;s										\\
		}
	}
	\caption{Structure of the local C-vines $\vine^s$, $\iN{s}{d}$.}
	\label{fig:cvines}
\end{figure}
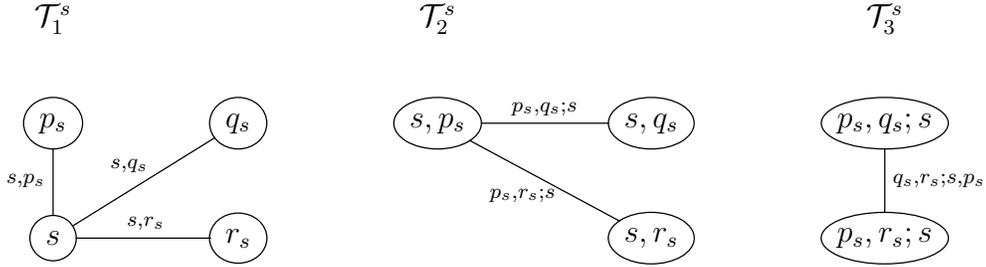

To model the dependence of the (transformed) data, we are going to set up a composite likelihood which is composed out of the vine copula densities corresponding to $d$ $4$-dimensional C-vines numbered $1,\ldots,d$, i.e. we consider one C-vine $\vine^s=(\tree_1^s,\tree_2^s,\tree_3^s)$ for every variable $\bu^s$, $\iN{s}{d}$. The first root node variable of every C-vine $\vine^s$ is labeled $s$. For each C-vine $\vine^s$ we select the variable $p_s$ with the shortest distance to the first root node variable $s$ as the second root variable, the variable $q_s$ with the second shortest distance as the third root variable and the variable $r_s$ with the third shortest distance as the forth variable of the C-vine. This results in root variable sets $\calR^s\coloneqq\left\{s,p_s,q_s,r_s\right\}$, $\iN{s}{d}$, which determine the C-vine collection $\cvc\coloneqq\left\{\vine^s: \iN{s}{d}\right\}$ uniquely, i.e. we have fixed the tree structure of all $d$ involved C-vines. We denote the respective edge sets by $\edge_1^s\coloneqq\left\{\{s,p_s\},\{s,q_s\},\{s,r_s\}\right\}$, $\edge_2^s\coloneqq\left\{\{p_s,q_s;s\},\{p_s,r_s;s\}\right\}$ and $\edge_3^s\coloneqq\left\{\{q_s,r_s;s,p_s\}\right\}$. Figure \ref{fig:cvines} is a graphical representation of C-vine $\vine^s$.

Adequate families for the bivariate copulas corresponding to the edges of $\vine^s$ have to be determined. The selection is conducted sequentially, based on the Akaike information criterion (AIC), where a specified set of different copula families is allowed. We are going to consider only one- or two-parametric families, where we denote the parameters by $\theta$ and $\nu$, respectively. If the family is one-parametric, we will ignore the second parameter $\nu$ $(\nu\coloneqq0)$.

Since now, all components of the local C-vine copulas are chosen, their likelihoods are given as
\begin{align*}
	\Lik{s}{\btheta_s, \bnu_s}{u^{s}_{t},u^{p_s}_{t},u^{q_s}_{t},u^{r_s}_{t}}	& =	\cop{s}{u^{s}_{t},u^{p_s}_{t},u^{q_s}_{t},u^{r_s}_{t} \sepi \btheta_s, \bnu_s}	\\
	& = \cop{s,p_s}{u^{s}_{t},u^{p_s}_{t} \sepi \te_{s,p_s},\nu_{s,p_s}} \cdot \cop{s,q_s}{u^{s}_{t},u^{q_s}_{t} \sepi \te_{s,q_s},\nu_{s,q_s}}	\\*
	& \cdot	\cop{s,r_s}{u^{s}_{t},u^{r_s}_{t} \sepi \te_{s,r_s},\nu_{s,r_s}}	\\*
	& \cdot	\cop{p_s,q_s;s}{u^{p_s|s}_{t},u^{q_s|s}_{t} \sepi \te_{p_s,q_s;s},\nu_{p_s,q_s;s}}	\\*
	& \cdot \cop{p_s,r_s;s}{u^{p_s|s}_{t},u^{r_s|s}_{t} \sepi \te_{p_s,r_s;s},\nu_{p_s,r_s;s}}	\\*
	& \cdot	\cop{q_s,r_s;s,p_s}{u^{q_s|s,p_s}_{t},u^{r_s|s,p_s}_{t} \sepi \te_{q_s,r_s;s,p_s},\nu_{q_s,r_s;s,p_s}}
\end{align*}
where the transformed variables are defined as
\beqo
	u^{o|s}_{t} \coloneqq	C_{o|s}(u^{o}_{t} \sepl u^{s}_{t}) =	\partialderiv{u^{s}_{t}} \Cop{o,s}{u^{o}_{t},u^{s}_{t} \sepi \te_{s,o},\nu_{s,o}}, \quad o=p_s,q_s,r_s,
\eeqo
\beqo
	u^{o|s,p_s}_{t} \coloneqq	C_{o|s,p_s}(u^{o}_{t} \sepl u^{s}_{t},u^{p_s}_{t}) = \partialderiv{u^{p_s|s}_{t}} \Cop{o,p_s;s}{u^{o|s}_{t},u^{p_s|s}_{t} \sepi \te_{p_s,o;s},\nu_{p_s,o;s}}, \quad o=q_s,r_s.
\eeqo
The corresponding parameters are summarized as
\begin{align*}
	\btheta_s	& \coloneqq	\left(\te_{s,p_s}, \te_{s,q_s}, \te_{s,r_s}, \te_{p_s,q_s;s}, \te_{p_s,r_s;s}, \te_{q_s,r_s;s,p_s}\right)\T,	\\
	\bnu_s	& \coloneqq	\left(\nu_{s,p_s}, \nu_{s,q_s}, \nu_{s,r_s}, \nu_{p_s,q_s;s}, \nu_{p_s,r_s;s}, \nu_{q_s,r_s;s,p_s}\right)\T.
\end{align*}
%Note that the second parameter $\nu_{ij|\ldots}$ of the bivariate copula is only needed in the case of a Student-$t$ copula, for the respective degrees of freedom. Otherwise the bivariate copula does not depend on $\nu_{ij|\ldots}$ and we omit this parameter or set it simply to zero.
Each first order tree edge $\left\{i,j\right\}$ can occur in up to two of the $d$ local C-vines, i.e. in $\vine^i$ and $\vine^j$. In this case we assume that $\te_{i,j}=\te_{j,i}$ and $\nu_{i,j}=\nu_{j,i}$, $\iN{i}{d}$, $\iN{j}{d}$. The edges of the higher order trees of the local C-vines $\vine^s$ are all unique, since their first conditioning variable is always the first root variable $s$ which differs for every $\vine^s$.

To be able to set up the composite likelihood, it remains to specify the weights $w_s$, $\iN{s}{d}$, corresponding to the likelihoods $\calL_{s}$ stated above. We define them as the reciprocal value of the number counts
\beqo
	n_s \coloneqq \#\left\{k: s \mbox{ is included in C-vine }\vine^k\right\},\ \iN{s}{d},
\eeqo
of local C-vines which include the variable $\iN{s}{d}$, i.e. $w_s \coloneqq n_s^{-1}$.

This results in the \emph{local C-vine composite likelihood} (LCVCL)
\beqo
	\Lik{\ast}{\btheta^\ast, \bnu^\ast}{\bu^1,\ldots,\bu^d}	=	\prodsd\prodtN\left[\Lik{s}{\btheta_s, \bnu_s}{u^{s}_{t},u^{p_s}_{t},u^{q_s}_{t},u^{r_s}_{t}}\right]^{w_s},
\eeqo
where $\btheta^\ast \coloneqq	\left\{\theta: \theta\in\btheta_s, \iN{s}{d}\right\}$ and $\bnu^\ast \coloneqq	\left\{\nu: \nu\in\bnu_s, \iN{s}{d}\right\}$.
Therefore the composite log-likelihood can be expressed as
\beqq\label{eq:compll}
	\cll{\ast}{\btheta^\ast, \bnu^\ast}{\bu^1,\ldots,\bu^d}	=	\sumsd\sumtN{w_s}\ll{s}{\btheta_s, \bnu_s}{u^{s}_{t},u^{p_s}_{t},u^{q_s}_{t},u^{r_s}_{t}},
\eeqq
where the respective local C-vine copula log-likelihoods are defined as
\beqo
	\ll{s}{\btheta_s, \bnu_s}{u^{s}_{t},u^{p_s}_{t},u^{q_s}_{t},u^{r_s}_{t}}	\coloneqq	\ln \Lik{s}{\btheta_s, \bnu_s}{u^{s}_{t},u^{p_s}_{t},u^{q_s}_{t},u^{r_s}_{t}}.
\eeqo

%%%%%%%%%%%%%%%%%%%%%%%%%%%%%%%%%%%%%%%% Section 03 %%%%%%%%%%%%%%%%%%%%%%%%%%%%%%%%%%%%%%%%

\subsection{LCVCL -- Application to mean temperature data}\label{sec:CVM:app}

%We apply the above composite likelihood approach to the mean temperature data set.
In a first step we have to select $54$ local C-vines $\vine^s$, $\iN{s}{54}$, as described in Section \ref{sec:CVM}. The respective choices are given in Table \ref{tab:CVM:roots}. Moreover we have to determine adequate families for the
%$6\cdot54=324$
bivariate copulas corresponding to the edges of the chosen $\vine^s$. For this we allow bivariate Gaussian ($\Phi$), Student-$t$ (t), Clayton (C), Gumbel (G) and Frank (F) copulas, as well as rotated versions of the Clayton and Gumbel copula. The exact type of rotation is fixed after the respective parameters are estimated. Using AIC we select the families as given in Table \ref{tab:CVM:fam}. This results in a parametrization with $464$ parameters.

\begin{table}[htb]
\centering
\caption[Root variables corresponding to the local C-vines $\vine^s$, $\iN{s}{54}$ (mean temperature data).]{Root variables corresponding to the local C-vines $\vine^s$, $\iN{s}{54}$, resulting from the application of the local C-vine composite likelihood approach to the mean temperature (training) data collected at $54$ observation stations.} 
\label{tab:CVM:roots}
\begin{tabular}{rrrr||rrrr||rrrr||rrrr}
	$s$ & $p_s$ & $q_s$ & $r_s$ & $s$ & $p_s$ & $q_s$ & $r_s$ & $s$ & $p_s$ & $q_s$ & $r_s$ & $s$ & $p_s$ & $q_s$ & $r_s$ \\ 
  \hline
	\hline
	1 & 4 & 49 & 40 & 15 & 42 & 43 & 37 & 29 & 35 & 24 & 41 & 43 & 53 & 39 & 41 \\ 
  2 & 24 & 10 & 29 & 16 & 33 & 6 & 50 & 30 & 38 & 11 & 7 & 44 & 52 & 32 & 49 \\ 
  3 & 17 & 31 & 36 & 17 & 3 & 13 & 36 & 31 & 3 & 38 & 26 & 45 & 25 & 48 & 34 \\ 
  4 & 1 & 40 & 23 & 18 & 10 & 8 & 23 & 32 & 21 & 50 & 44 & 46 & 9 & 21 & 32 \\ 
  5 & 36 & 12 & 3 & 19 & 26 & 6 & 35 & 33 & 16 & 20 & 6 & 47 & 51 & 34 & 54 \\ 
  6 & 22 & 16 & 50 & 20 & 33 & 24 & 23 & 34 & 13 & 25 & 47 & 48 & 45 & 25 & 36 \\ 
  7 & 22 & 9 & 21 & 21 & 32 & 50 & 9 & 35 & 29 & 54 & 24 & 49 & 1 & 52 & 40 \\ 
  8 & 23 & 18 & 10 & 22 & 6 & 50 & 7 & 36 & 5 & 17 & 13 & 50 & 6 & 16 & 22 \\ 
  9 & 21 & 46 & 7 & 23 & 8 & 20 & 10 & 37 & 39 & 27 & 43 & 51 & 47 & 27 & 28 \\ 
  10 & 18 & 8 & 2 & 24 & 20 & 29 & 2 & 38 & 30 & 31 & 12 & 52 & 40 & 44 & 49 \\ 
  11 & 30 & 7 & 9 & 25 & 45 & 34 & 48 & 39 & 37 & 43 & 41 & 53 & 43 & 41 & 29 \\ 
  12 & 5 & 38 & 31 & 26 & 19 & 31 & 3 & 40 & 4 & 52 & 1 & 54 & 41 & 35 & 13 \\ 
  13 & 34 & 17 & 36 & 27 & 51 & 37 & 39 & 41 & 53 & 54 & 29 &  &  &  &  \\ 
	14 & 28 & 47 & 45 & 28 & 51 & 27 & 14 & 42 & 15 & 37 & 43 &  &  &  &  \\ 
\end{tabular}
\end{table}

\begin{table}[htb]
\centering
\caption[Families corresponding to the C-vine edges $\edge_1^s$, $\edge_2^s$ and $\edge_3^s$, $\iN{s}{54}$ (mean temperature data).]{Families corresponding to the C-vine edges $\edge_1^s$, $\edge_2^s$ and $\edge_3^s$, $\iN{s}{54}$, resulting from the application of the local C-vine composite likelihood method to the mean temperature (training) data collected at $54$ observation stations.} 
\label{tab:CVM:fam}
\begin{tabular}{p{9pt}|C{5pt}C{5pt}C{5pt}C{5pt}C{5pt}C{5pt}||p{9pt}|C{5pt}C{5pt}C{5pt}C{5pt}C{5pt}C{5pt}||p{9pt}|C{5pt}C{5pt}C{5pt}C{5pt}C{5pt}C{5pt}}
	$s$ & \rotatebox{90}{$sp_s$} & \rotatebox{90}{$sq_s$} & \rotatebox{90}{$sr_s$} & \rotatebox{90}{$p_sq_s;s$} & \rotatebox{90}{$p_sr_s;s$} & \rotatebox{90}{$q_sr_s;sp_s$} & $s$ & \rotatebox{90}{$sp_s$} & \rotatebox{90}{$sq_s$} & \rotatebox{90}{$sr_s$} & \rotatebox{90}{$p_sq_s;s$} & \rotatebox{90}{$p_sr_s;s$} & \rotatebox{90}{$q_sr_s;sp_s$} & $s$ & \rotatebox{90}{$sp_s$} & \rotatebox{90}{$sq_s$} & \rotatebox{90}{$sr_s$} & \rotatebox{90}{$p_sq_s;s$} & \rotatebox{90}{$p_sr_s;s$} & \rotatebox{90}{$q_sr_s;sp_s$} \\ 
  \hline
	\hline
	1 & t & t & t & t & t & t & 19 & t & t & t & t & t & C & 37 & t & t & t & t & t & t \\ 
  2 & t & t & t & t & t & t & 20 & t & t & t & t & t & G & 38 & t & t & t & t & t & t \\ 
  3 & t & t & t & t & t & C & 21 & t & t & t & t & t & C & 39 & t & t & t & t & t & t \\ 
  4 & t & t & t & t & t & G & 22 & t & t & t & t & t & t & 40 & t & t & t & C & t & t \\ 
  5 & t & t & t & t & t & t & 23 & t & t & t & t & t & t & 41 & t & t & t & t & t & G \\ 
  6 & t & t & t & F & t & t & 24 & t & t & t & t & t & t & 42 & t & t & t & t & t & $\Phi$ \\ 
  7 & t & t & t & C & F & t & 25 & t & t & t & t & t & t & 43 & t & t & t & G & t & t \\ 
  8 & t & t & t & G & t & t & 26 & t & t & t & t & t & t & 44 & t & G & t & t & t & t \\ 
  9 & t & t & t & t & t & t & 27 & t & t & t & t & t & t & 45 & t & t & t & F & t & t \\ 
  10 & t & t & t & t & t & t & 28 & t & t & t & t & t & G & 46 & t & t & G & t & G & t \\ 
  11 & t & t & t & t & F & t & 29 & t & t & t & F & t & t & 47 & t & t & t & C & t & t \\ 
  12 & t & t & t & t & t & t & 30 & t & t & t & t & t & t & 48 & t & t & t & t & t & t \\ 
  13 & t & t & t & t & G & t & 31 & t & t & t & t & t & t & 49 & t & t & t & t & t & t \\ 
  14 & t & t & t & t & t & t & 32 & t & t & G & t & t & G & 50 & t & t & t & t & t & t \\ 
  15 & t & t & G & $\Phi$ & t & $\Phi$ & 33 & t & t & t & t & t & C & 51 & t & t & t & t & t & t \\ 
  16 & t & t & t & t & t & t & 34 & t & t & t & F & t & t & 52 & t & t & t & t & t & t \\ 
  17 & t & t & t & t & t & t & 35 & t & t & t & t & t & t & 53 & t & t & t & t & t & t \\ 
  18 & t & t & t & t & t & t & 36 & t & t & t & G & t & t & 54 & t & t & t & t & t & t \\ 
\end{tabular}
\end{table}

\begin{figure}[htb]
	\centering
		\includegraphics[width=0.48\textwidth]{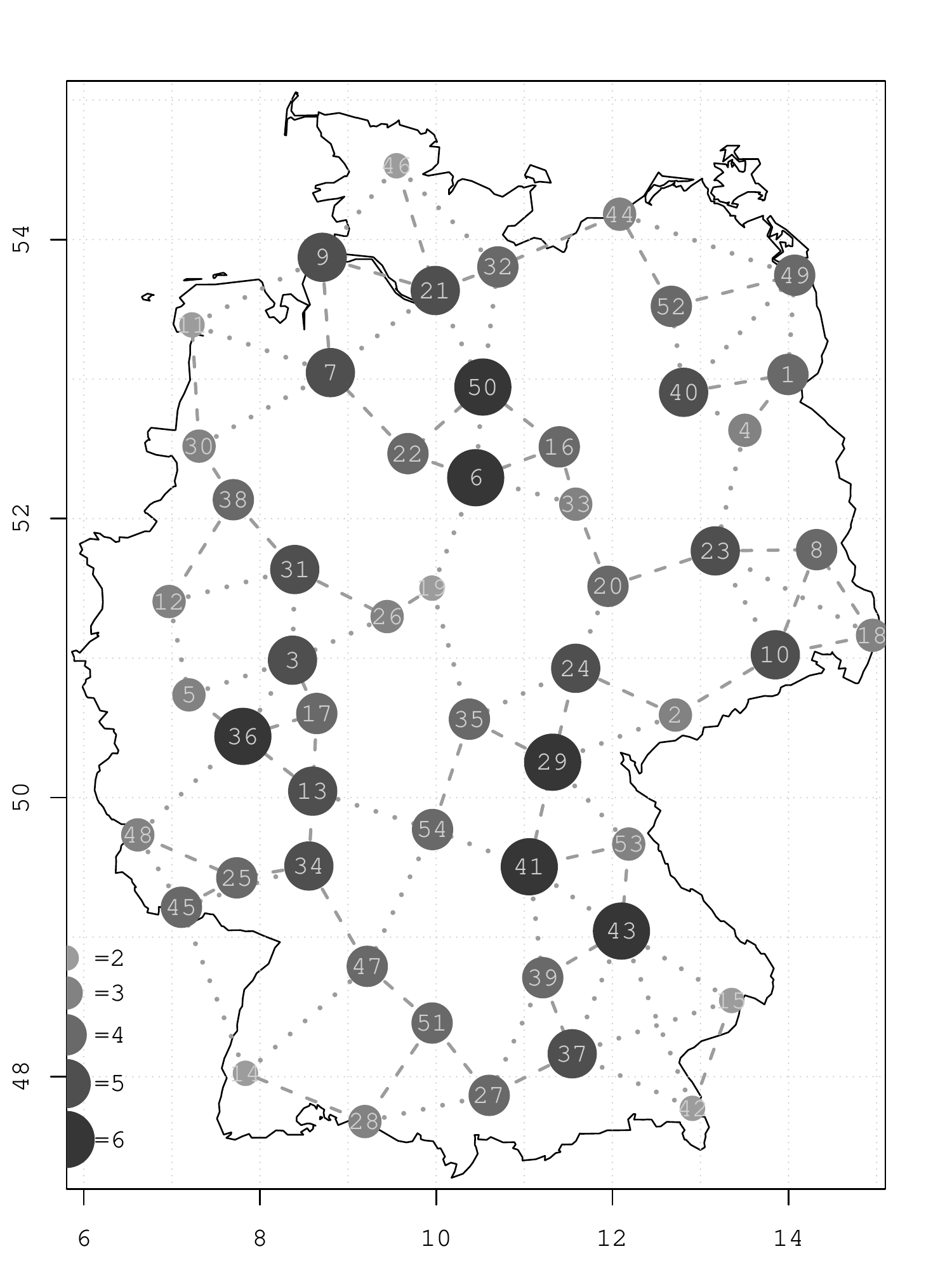}
	\caption[LCVCL structure]{LCVCL structure: Visualization of the spatial structure of a local C-vine composite likelihood for the training data set.}
	\label{fig:GermanySCVM}
\end{figure}

Figure \ref{fig:GermanySCVM} illustrates the chosen structure. The numbered circles indicate the locations of the $54$ observation stations. Their circumference and color represent the numbers $n_s$ of C-vines in which the location $s$ is included. Further the lines indicate all local C-vine edges $\edge_1^s$ that occur in the first trees of the local C-vines $\vine^s$. Dotted lines represent edges that occur only once, whereas dashed lines represent edges that occur in the first tree of two local C-vines.

Applied to the mean temperature data set numerical maximization of the composite log-likelihood \eqref{eq:compll} yields a maximum composite log-likelihood of $\cll{\ast}{\wh\btheta^\ast, \wh\bnu^\ast}{\bu^1,\ldots,\bu^d} = 42106.41$. %The respective AIC and BIC are given by $\AIC_\ast = -83222.8$ and $\BIC_\ast = -80749.5$.
Note that the estimated degrees of freedom parameters for Student-$t$ copulas range between $3.9$ and $25.8$.

%%%%%%%%%%%%%%%%%%%%%%%%%%%%%%%%%%%%%%%% Section 04 %%%%%%%%%%%%%%%%%%%%%%%%%%%%%%%%%%%%%%%%

\subsection{Spatial local C-vine composite likelihood (S-LCVCL)}\label{sec:SCVM}

In a second step we utilize spatial information to reduce the number of parameters in the local C-vine composite likelihood approach. For this we consider regressions of the first copula parameters $\btheta^\ast$ on spatial covariates. Furthermore we are interested in a joint modeling of the second copula parameters $\bnu^\ast$.

Let us now consider an arbitrary local C-vine $\vine^s$ of a C-vine collection $\cvc$, with root variable set $\calR^s\coloneqq\left\{s,p_s,q_s,r_s\right\}$, and the respective vine copula. Let furthermore $\{i,j;\calD\}$ denote an arbitrary edge in $\edge_1^s\cup\edge_2^s\cup\edge_3^s$, where $\calD$ may be the empty set. As already outlined in \cite{erhardt14} we are able to transform between a first copula parameter $\te_{i,j;\calD}$ and its respective Kendall's $\tau_{i,j;\calD}$, applying some known (see e.g. Tables 1 and 2 of \cite{cdvine13}), copula family ($b_{i,j;\calD}$) dependent transformation
\beqo
	\tau_{i,j;\calD} = g_\tau(\te_{i,j;\calD};b_{i,j;\calD}).
\eeqo
Since $\tau_{i,j;\calD}\in[-1,1]$ we use the Fisher z-transformation defined as
\beqo
	\xi \coloneqq g_z(r) = \half \ln\left(\frac{1+r}{1-r}\right),\, r\in(-1,1),
\eeqo
to obtain response variables $\xi \in \R$. On the scale of $\xi$ we consider tree-specific ($l=1,2,3$) regression models
\beqq\label{eq:SCVM:linmod}
	\xi_{i,j;\calD} \coloneqq g_z(\tau_{i,j;\calD}) = h_{l}\left(i,j,\calD|\bbeta_{l}\right) + \veps_{i,j;\calD}, \quad \veps_{i,j;\calD}\simiid\vN{0}{\sigma^2}.
\eeqq
The functions $h_{l}$, $l=1,2,3$, are linear, tree-dependent model functions of the form
\beqo
	h_{l}\left(i,j,\calD|\bbeta_{l}\right) = \beta_{0;l} + \sum_{v=1}^{n_v^l}\left(\beta^v_{i,j;l}w^v_{i,j} + \sum_{k\in\calD}\beta^v_{i,k;l}w^v_{i,k} + \sum_{k\in\calD}\beta^v_{j,k;l}w^v_{j,k}\right),
\eeqo
where $w^v_{p,q}$, $v=1,\ldots,n_v^l$, are $n_v^l$ different types of spatial covariates measuring a specific spatial difference between the locations $p$ and $q$. The corresponding parameters are collected in $\bbeta_{l}\in\R^{1+n_v^l(1+2(l-1))}$, $l=1,2,3$.
%which consider linear relationships of the Fisher z-transformed rank correlations $\tau_{i,j;\calD}$ on the logarithmized station distances $d_{i,j}$, $d_{i,k}$, $d_{j,k}$ and elevation differences $e_{i,j}$, $e_{i,k}$, $e_{j,k}$, where $k\in\calD$.
%For an extensive tree-wise analysis of the above stated regression models we refer the reader to Section 6.4 of \cite{erhardt13}, where different model specifications for the first copula parameters $\te_{i,j;\calD}$ were investigated.
This leads to model specifications for $\te_{i,j;\calD}$ of the form
\beqq\label{eq:SCVM:repar}
	\te_{i,j;\calD}	\coloneqq \itau\left(\Fzi \left(h_{l}(i,j,\calD|\bbeta_{l})\right);b_{i,j;\calD}\right).
\eeqq

We follow a similar regression approach to specify the second copula parameters $\nu_{i,j;\calD}\in(0,\infty)$. Model specifications
\beqq\label{eq:SCVM:nu}
	\nu_{i,j;\calD}	\coloneqq \exp\left(h_\nu(i,j,\calD,l|\bbeta_\nu)\right),
\eeqq
with linear model functions $h_\nu$ and parameter vector $\bbeta_\nu$ are considered. %\in \R^{n_\text{par}^\nu}$.

A local C-vine composite likelihood using specifications \eqref{eq:SCVM:repar} and \eqref{eq:SCVM:nu}, will be called \emph{spatial local C-vine composite likelihood} (S-LCVCL). In order to estimate the respective model parameters, the composite log-likelihood \eqref{eq:compll} parametrized via \eqref{eq:SCVM:repar} and \eqref{eq:SCVM:nu} has to be maximized.
%This yields the maximum composite likelihood estimates (mcle)
%\beqo
	%\wh\bbeta_{\text{mcle}}^{\ast} = \left(\wh\bbeta_{\text{mod}}^{\ast}, \wh\bbeta_\nu^{\ast}\right)\T \in \R^{(n_\text{par}^\text{mod}+n_\text{par}^\nu)}.
%\eeqo

%%%%%%%%%%%%%%%%%%%%%%%%%%%%%%%%%%%%%%%% Section 05 %%%%%%%%%%%%%%%%%%%%%%%%%%%%%%%%%%%%%%%%

\subsection{S-LCVCL -- Application to mean temperature data}\label{sec:SCVM:app}

%Now we are going to apply the spatial local C-vine composite likelihood approach to the mean temperature data.
We have to choose suitable model specifications \eqref{eq:SCVM:repar} and \eqref{eq:SCVM:nu}. This selection is performed based on an analysis of the results in Section \ref{sec:CVM:app} from fitting a LCVCL to the mean temperature data. The spatial covariates which will be taken into consideration are logarithmized spatial distances $D_{p,q}$ and elevation differences $E_{p,q}$ between station pairs $(p,q)$, $p=1,\ldots,d$, $p\neq q=1,\ldots,d$.

For a detailed analysis and comparison of different model specifications, using different combinations of the spatial covariates $\lnf{D_{i,j}}$, $\lnf{D_{i,k}}$, $\lnf{D_{j,k}}$, $\lnf{E_{i,j}}$, $\lnf{E_{i,k}}$, $\lnf{E_{j,k}}$, where $k\in\calD$,  we refer to Sections 6.4 and 6.5 of \cite{erhardt13}. There the explanatory power of these model specifications is compared by tree-wise ($l=1,2,3$) analysis and comparison of linear models of the form \eqref{eq:SCVM:linmod} and the respective coefficients of determination ($\Rsq$ respectively $\Rsqa$).
%Moreover the required number of parameters was of interest for the selection of a model specification.

The investigations in \cite{erhardt13} show, that the distance based regressors have high explanatory power, whereas the elevation difference based regressors perform comparatively bad. Only in the first trees the contribution of the elevation difference seems to be reasonable. Therefore we choose the model specification \eqref{eq:SCVM:repar} according to
\begin{align*}
	h_{1}\left(i,j,\emptyset|\bbeta_{1}\right) \coloneqq \beta_{0;1}	+ \beta^1_{i,j;1}\lnf{D_{i,j}} &	+ \beta^2_{i,j;1}\lnf{E_{i,j}}	\\
	h_{2}\left(i,j,\{k\}|\bbeta_{2}\right) \coloneqq \beta_{0;2}	+ \beta^1_{i,j;2}\lnf{D_{i,j}}	&	+ \beta^1_{i,k;2}\lnf{D_{i,k}}	+ \beta^1_{j,k;2}\lnf{D_{j,k}}	\\
	h_{3}\left(i,j,\{k,m\}|\bbeta_{3}\right) \coloneqq \beta_{0;3}	+ \beta^1_{i,j;3}\lnf{D_{i,j}}	&	+ \beta^1_{i,k;3}\lnf{D_{i,k}}	+ \beta^1_{i,m;3}\lnf{D_{i,m}}	\\*
				&	+ \beta^1_{j,k;3}\lnf{D_{j,k}}	+ \beta^1_{j,m;3}\lnf{D_{j,m}},
\end{align*}
with %respective parameter vectors
$\bbeta_{1}	\coloneqq \left(\beta_{0;1}, \beta^1_{i,j;1}, \beta^2_{i,j;1}\right)\T \in \R^{3}$,
$\bbeta_{2}	\coloneqq \left(\beta_{0;2}, \beta^1_{i,j;2}, \beta^1_{i,k;2}, \beta^1_{j,k;2}\right)\T \in \R^{4}$,
$\bbeta_{3}	\coloneqq \left(\beta_{0;3}, \beta^1_{i,j;3}, \beta^1_{i,k;3}, \beta^1_{i,m;3}, \beta^1_{j,k;3}, \beta^1_{j,m;3}\right)\T \in \R^{6}$,
summarized in the vector $\bbeta_\te^{\ast} \coloneqq \left(\bbeta_{1}\T,\bbeta_{2}\T,\bbeta_{3}\T\right)\T \in \R^{13}$.

Similar investigations were performed for the second copula parameter model specification \eqref{eq:SCVM:nu}. Regression of the parameters $\lnf{\nu_{i,j;\calD}}$ on the tree number $l$, the log-distances $\lnf{D_{i,j}}$ and log-elevation differences $\lnf{E_{i,j}}$ showed a linear relationship with the tree number and the log-distances. Hence we define the respective model function as
\beqo
	h_\nu(i,j,\calD,l|\bbeta_\nu^{\ast}) \coloneqq \beta_{0}^{\nu} + \beta_{1}^{\nu}l + \beta_{2}^{\nu}\lnf{D_{i,j}},
\eeqo
with $\bbeta_\nu^{\ast} \coloneqq (\beta_{0}^{\nu},\beta_{1}^{\nu},\beta_{2}^{\nu})\T \in \R^3$, resulting in the linear regression model
\beqq\label{eq:SCVM:linmodnu}
	\lnf{\nu_{i,j;\calD}} = h_\nu(i,j,\calD,l|\bbeta_\nu^{\ast}) + \veps^{\nu}_{i,j;\calD}, \quad \veps^{\nu}_{i,j;\calD}\simiid\vN{0}{\sigma^2}.
\eeqq

Next we fit the selected spatial local C-vine composite likelihood to the data. This is conducted by maximization of the composite log-likelihood \eqref{eq:compll} which is parametrized by \eqref{eq:SCVM:repar} and \eqref{eq:SCVM:nu}, where the model functions $h_{l}$, $l=1,2,3$ and $h_\nu$ are specified as above. Suitable start values for the numerical optimization procedure  are obtained from fitting the linear models \eqref{eq:SCVM:linmod} and \eqref{eq:SCVM:linmodnu} to the estimates $\hattau^\ast \coloneqq g_\tau\left(\hattheta^\ast\right)\coloneqq\left(g_\tau\left(\wh\te\right)\right)_{\wh\te\in\hattheta^\ast}$ and $\hatnu^\ast$ obtained in Section \ref{sec:CVM:app}.

This time we get %$\cll{\ast}{\wh\bbeta_\te^{\ast}, \wh\bbeta_\nu^{\ast}}{\bu^1,\ldots,\bu^d}=$
$40938.80$ for the maximum composite log-likelihood.
%The respective AIC and BIC are given by $\AIC_\ast=-81845.6$ and $\BIC_\ast=-81765.7$.
%Compared to the composite vine model the composite log-likelihood of the spatial composite vine model shrinked by about $1168$.
%Due to the fact that the number of parameters is reduced considerably, the BIC of the spatial composite vine model is smaller. Comparison based on the AIC would however favor the composite vine model.
The fitting procedure yields the parameter vector estimate $\wh\bbeta^{\ast} \coloneqq (\wh\bbeta_\te^{\ast}, \wh\bbeta_\nu^{\ast})\T \in \R^{16}$. Its components are given in Table \ref{tab:SCVMmcle} together with the respective start values.

% latex table generated in R 2.15.0 by xtable 1.7-1 package
% Tue Jul 16 21:45:17 2013
\begin{table}[htb]
\centering
\caption[Start values and maximum composite-likelihood estimates of the spatial local C-vine composite likelihood for the mean temperature data.]{Start values (start) and maximum composite-likelihood estimates (mcle) of the spatial local C-vine composite likelihood for the mean temperature data.} 
\label{tab:SCVMmcle}
%\footnotesize
\begin{tabular}{rlrr|rlrr|rlrr}
	no. & par & start & mcle & no. & par & start & mcle & no. & par & start & mcle \\ 
  \hline
	\hline
	1 & $\beta_{0;1}$ & 2.30 & 2.25 & 7 & $\beta^1_{j,k;2}$ & 0.30 & 0.25 & 13 & $\beta^1_{j,m;3}$ & 0.12 & 0.12 \\ 
  2 & $\beta^1_{i,j;1}$ & -0.31 & -0.30 & 8 & $\beta_{0;3}$ & 0.26 & 0.70 & 14 & $\beta_0^\nu$ & 0.68 & 0.40 \\ 
  3 & $\beta^2_{i,j;1}$ & -0.03 & -0.02 & 9 & $\beta^1_{i,j;3}$ & -0.43 & -0.41 & 15 & $\beta_1^\nu$ & 0.15 & 0.02 \\ 
  4 & $\beta_{0;2}$ & 0.31 & 0.58 & 10 & $\beta^1_{i,k;3}$ & 0.11 & 0.08 & 16 & $\beta_2^\nu$ & 0.26 & 0.27 \\ 
  5 & $\beta^1_{i,j;2}$ & -0.51 & -0.48 & 11 & $\beta^1_{i,m;3}$ & 0.09 & 0.02 &  &  &  &  \\ 
  6 & $\beta^1_{i,k;2}$ & 0.22 & 0.18 & 12 & $\beta^1_{j,k;3}$ & 0.12 & 0.10 &  &  &  &  \\ 
\end{tabular}
\end{table}

%%%%%%%%%%%%%%%%%%%%%%%%%%%%%%%%%%%%%%%% Section 05 %%%%%%%%%%%%%%%%%%%%%%%%%%%%%%%%%%%%%%%%

\subsection{Spatial local C-vine based prediction}\label{sec:app:SCVMpred}

Our next goal is a local C-vine based prediction method for unobserved locations, which exploits the available spatial information. Our previous modeling approach using local C-vines suggests to perform prediction based on a four-dimensional C-vine copula which is composed out of the variables corresponding to the new location $s$ from which to predict and to its three closest neighbors among the stations $1,\ldots,d$. As before we denote the closest observation station as $p_s$, the second closest neighboring station as $q_s$ and the third closest station as $r_s$.
%\beqo
	%\wh u_s = F_{s|p_sq_sr_s}^{-1}(v_s|u_{p_s},u_{q_s},u_{r_s}), \quad \mbox{for } v_s\sim\vU01
%\eeqo
%The values of $u_{p_s}$, $u_{q_s}$ and $u_{r_s}$ are known (training data set on which the model is built).

\begin{figure}[htb]
	\centerline{
		\entrymodifiers={++[o][F]}
		%\SelectTips{cm}{}
		\xymatrixrowsep{0pc}\xymatrix @+2pc {
		p_s									& q_s \ar@{-}[l]_{p_s,q_s}		&	p_s,q_s											& *\txt{}													& q_s,r_s;p_s \ar@{-}[d]^{r_s,s;p_s,q_s}	\\
		s	\ar@{-}[u]_{p_s,s}	& r_s \ar@{-}[ul]_{p_s,r_s}	&	p_s,s \ar@{-}[u]_{q_s,s;p_s}	& p_s,r_s \ar@{-}[ul]_{q_s,r_s;p_s}	& q_s,s;p_s										\\
		}
	}
	\caption{C-vine structure used to sample from $C_{s|p_s,q_s,r_s}$.}
	\label{fig:pred:cvine}
\end{figure}
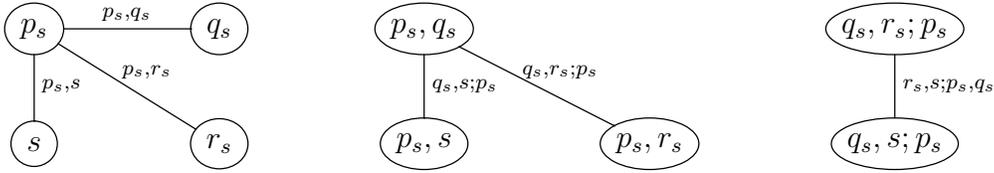

To predict we are going to sample from the conditional distribution $C_{s|p_s,q_s,r_s}(u^s|u^{p_s},u^{q_s},u^{r_s})$. It has to be calculated iteratively in analogy to Equation \eqref{eq:joe}. The C-vine utilized for the prediction is set up based on the pair-copulas needed for these calculations. For details on the technical derivation, which yields a C-vine as illustrated in Figure \ref{fig:pred:cvine}, we refer to Subsection 6.5.3 of \cite{erhardt13}. The respective C-vine copula will be called \emph{predictive C-vine copula} in the following. Note that the predictive C-vine differs from the C-vines $\vine^s$ defining the components of the (spatial) local C-vine composite likelihood (compare Figure \ref{fig:cvines}).

%\begin{figure}[htb]
	%\centerline{
		%\entrymodifiers={++[o][F]}
		%%\SelectTips{cm}{}
		%\xymatrix @+2pc {
		%*{\tree_1}					& *\txt{}										&	*{\tree_2}									& *\txt{}													&	*{\tree_3}							\\
		%p_s									& q_s \ar@{-}[l]_{p_sq_s}		&	p_sq_s											& *\txt{}													& q_sr_s|p_s \ar@{-}[d]^{r_ss|p_sq_s}	\\
		%s	\ar@{-}[u]_{p_ss}	& r_s \ar@{-}[ul]_{p_sr_s}	&	p_ss \ar@{-}[u]_{q_ss|p_s}	& p_sr_s \ar@{-}[ul]_{q_sr_s|p_s}	& q_ss|p_s										\\
		%}
	%}
	%\caption{C-vine structure in order to sample from $F_{s|p_sq_sr_s}(u^s|u^{p_s},u^{q_s},u^{r_s})$.}
	%\label{fig:pred:cvine}
%\end{figure}

In order to be able to simulate from $C_{s|p_s,q_s,r_s}$, we have to specify the copula families $b_{i,j;\calD}$ and the parameters $\te_{i,j;\calD}$ of the predictive C-vine copula. The first copula parameters are estimated according to Equation \eqref{eq:SCVM:repar} as
\beqo
	\wh\te_{i,j;\calD}	= \itau\left(\Fzi \left(h_{l}(i,j,\calD|\wh\bbeta_{l})\right);b_{i,j;\calD}\right).
\eeqo
using the model functions $h_{l}$, $l=1,2,3$, corresponding to the selected model specification and the parameter estimates $\wh\bbeta_{l}$ obtained from maximizing the spatial local C-vine composite likelihood. For the mean temperature data most pair-copulas of the local C-vine copulas were determined to be Student-$t$ copulas (see Table \ref{tab:CVM:fam}). Hence we choose the Student-$t$ family for all families $b_{i,j;\calD}$ of the predictive C-vine copula. Moreover we have to determine a second copula parameter for each of the six Student-$t$ copulas involved. According to Equation \eqref{eq:SCVM:nu} and using the maximum composite likelihood estimates $\wh\bbeta_\nu$, we obtain
\beqo
	\wh\nu_{i,j;\calD}	= \exp\left(h_\nu(i,j,\calD,l|\wh\bbeta_\nu)\right).
\eeqo

Furthermore we are interested in the predictive density $c_{s|p_s,q_s,r_s}$ corresponding to $C_{s|p_s,q_s,r_s}$. It is easily derived from the C-vine copula density given in Equation \eqref{eq:cvdensity}. We obtain
\begin{align*}%\label{eq:SCVMpdist}
	c_{s|p_s,q_s,r_s}(u^s|u^{p_s},u^{q_s},u^{r_s})	&	= \cop{p_s,s}{u^{p_s},u^{s}\sepi\wh\te_{p_s,s},\wh\nu_{p_s,s}}	\nonumber	\\*	
		&	\cdot	\cop{q_s,s;p_s}{u^{q_s|p_s},u^{s|p_s}\sepi\wh\te_{q_s,s;p_s},\wh\nu_{q_s,s;p_s}}	\\*
		&	\cdot	\cop{r_s,s;p_s,q_s}{u^{r_s|p_s,q_s},u^{s|p_s,q_s}\sepi\wh\te_{r_s,s;p_s,q_s},\wh\nu_{r_s,s;p_s,q_s}},	\nonumber
\end{align*}
where the transformed variables are obtained as $u^{o|p_s} \coloneqq	C_{o|p_s}(u^{o} \sepl u^{p_s})$, $o=s,q_s,$ and $u^{o|p_s,q_s} \coloneqq C_{o|p_s,q_s}(u^{o} \sepl u^{p_s},u^{q_s})$, $o=s,r_s$.
%\beqo
	%\breve{u}^{o} =	C(u^{o} \sepl u^{p_s}) =	\partialderiv{u^{p_s}} \Cop{op_s}{u^{o},u^{p_s} \sepi \te_{p_so},\nu_{p_so}}, \quad o=s,q_s,
%\eeqo
%\beqo
	%\check{u}^{o} =	C(u^{o} \sepl u^{p_s},u^{q_s}) = \partialderiv{\breve{u}^{q_s}} \Cop{oq_s;p_s}{\breve{u}^{o},\breve{u}^{q_s} \sepi \te_{q_so|p_s},\nu_{q_so|p_s}}, \quad o=s,r_s.
%\eeqo

This concludes the required methodology for local C-vine based sampling on the copula data level. Hence we are able to simulate copula data $u_t^s$ for an arbitrary location $s$ and and arbitrary point in time $t$, given the respective copula data $u_t^{p_s}$, $u_t^{q_s}$, $u_t^{r_s}$ corresponding to the three closest neighboring stations. However the copula data obtained from these simulations has to be transformed back to the original data level to be able to perform prediction. Using the notation introduced in Section \ref{sec:data}, the back transformation of the simulated copula data is performed by calculating $y_t^s=g(t, \bx^s; \wh\bbeta)+(F^s)^{-1}(u_t^s)$.

A detailed explanation of the back transformation in the case of the mean temperature data is given in the Web Supplementary Materials of \cite{erhardt14}.
The temperatures resulting from a sufficiently high number of simulations can then be used to calculate temperature predictions.

%% file: Sections/Validation.tex
\section[Model validation and comparison]{Model validation and comparison}\label{sec:appl}

In this section we compare our spatial local C-vine composite likelihood approach (SC) to the spatial R-vine model (SV) and the spatial Gaussian model (SG), which were presented in \cite{erhardt14}. To be able to evaluate the quality of the predictions, we predict at the locations given in the validation data set. Each prediction is based on $1000$ simulations from the particular predictive distribution.

%%%%%%%%%%%%%%%%%%%%%%%%%%%%%%%%%%%%%%%% Section 01 %%%%%%%%%%%%%%%%%%%%%%%%%%%%%%%%%%%%%%%%

%\subsection{Spatial model comparison}\label{sec:spatmodcomp}

\begin{figure}[htbp]
	\centering
		\includegraphics[width=0.80\textwidth]{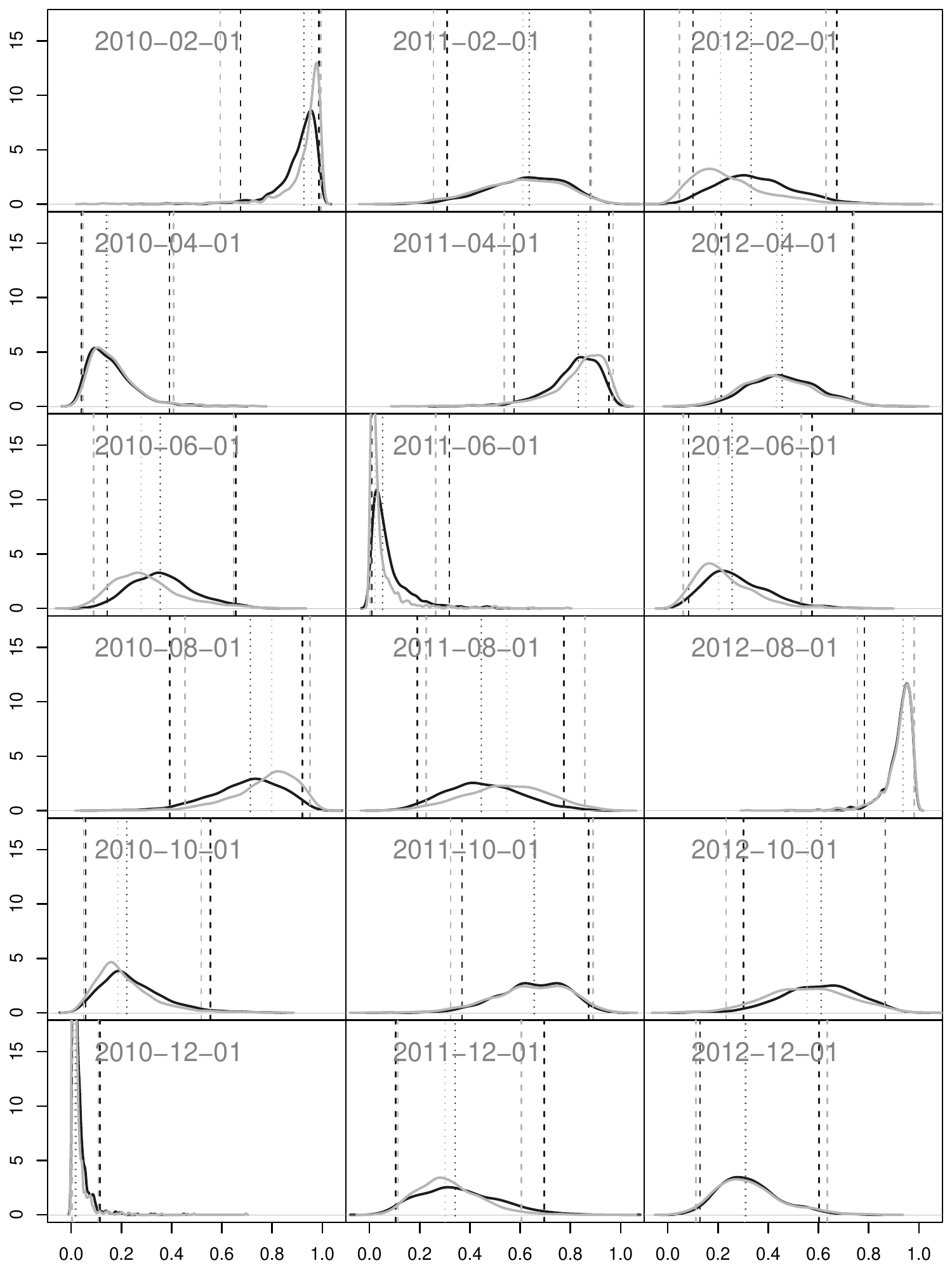}
	\caption[Comparison of kernel density estimates, calculated based on the predictive distributions arising from the spatial R-vine model and the spatial local C-vine composite likelihood approach, for the first of February, April, June, August, October and December $2010$-$2012$ at the observation station Arkona (56).]{Comparison of kernel density estimates, each calculated based on $1000$ simulations from the predictive distributions arising from the spatial R-vine model (black) and the spatial local C-vine composite likelihood approach (gray), for the first of February, April, June, August, October and December $2010$-$2012$ at the observation station Arkona (56). The respective medians and $95\%$ prediction intervals are indicated as vertical lines in the corresponding color.}
	\label{fig:PredDistrArko}
\end{figure}

First we compare the densities of the predictive distributions of the spatial R-vine model and the spatial local C-vine composite likelihood approach. Kernel density estimates of these densities, for the first of February, April, June, August, October and December $2010$-$2012$ at the observation station \emph{Arkona} (56) are illustrated in Figure \ref{fig:PredDistrArko}. The corresponding medians and $95\%$ prediction intervals are indicated.
%We do not observe any kind of temporal dependence, which is in agreement with the modeling assumption that the marginal model should capture all temporal dependencies.
%On the one hand there are some days like the first of August $2012$, where both densities are nearly identical, contrariwise there are many days like the first of February $2012$, where the densities differ considerably, however 
All prediction densities behave quite similar in terms of shape, scale and skew.

%\paragraph{Score based model comparison}\label{sec:scoremodcomp}

\begin{figure}[htb]
	\centering
		\includegraphics[width=1.00\textwidth]{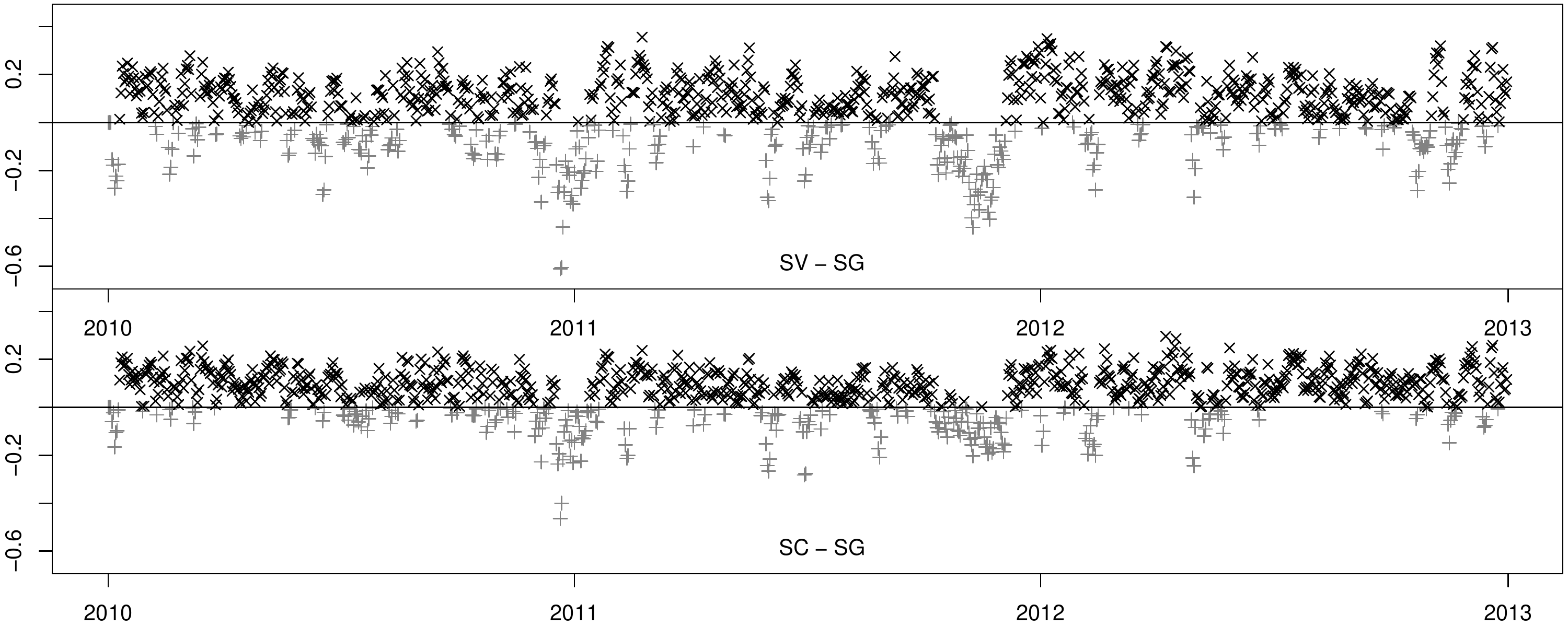}
	\caption[Log-score difference plots of the averaged continuous ranked probability scores comparing the different approaches (averaged over all $19$ observation stations of the validation data set).]{Log-score difference plots of the averaged continuous ranked probability scores comparing the spatial R-vine model and the spatial local C-vine composite likelihood approach to the corresponding averaged spatial Gaussian model scores (average over all $19$ observation stations of the validation data set). Points in time where the first approach has the lower average scores are marked by a black x. On the other hand, points in time where the second approach has the lower average scores are marked by a gray plus sign.}
	\label{fig:AverageScoresOP}
\end{figure}

Further comparisons are based on (negatively oriented) \textit{continuous ranked probability scores} (CRPS) (see Section 4.2 of \cite{gneiting07}). They assess the quality of our predictions.
%Negatively oriented means that smaller scores, i.e. scores closer to zero indicate a better fit.
%The scores will allow for an adequate comparative model validation.
The investigations are divided into three parts. First we compare averaged continuous ranked probability scores (Table \ref{tab:scores}, $\ov{\text{CRPS}}$). Afterwards percentaged model outperformance (Table \ref{tab:scores}, $\%$) and log-score difference plots (Figure \ref{fig:AverageScoresOP}) as introduced in \cite{erhardt14} are considered.

\paragraph{Averaged scores}
The averaged continuous ranked probability scores (CRPS) corresponding to the predictions from our three different approaches are given in Table \ref{tab:scores}, where we average over time. The overall averages are provided in addition. Scores close to zero are preferred. Thus the averaged scores in Table \ref{tab:scores} give a preference to the spatial local C-vine composite likelihood approach over the spatial R-vine model and over the spatial Gaussian model.

% latex table generated in R 3.0.2 by xtable 1.7-1 package
% Mon Jan 20 12:03:27 2014
\begin{table}[htbp]
\centering
\caption[Comparison of averaged CRPS and percentaged outperformance.]{Comparison of the averaged CRPS ($\ov{\text{CRPS}}$) of the spatial R-vine model (SV), the spatial local C-vine composite likelihood approach (SC) and the spatial Gaussian model (SG) and percentaged outperformance ($\%$) in terms of CRPS over the period $01/01/2010-12/31/2012$ for the observation stations of the validation data set. Here we define $A_1 \stackrel{\%}{\succ} A_2$ as the share of the points in time for which Approach $A_1$ is preferred over Approach $A_2$ in terms of CRPS.} 
\label{tab:scores}
\begin{tabular}{rl|ccc|ccc}
  \hline
	\hline
 	& short & \multicolumn{3}{c|}{$\ov{\text{CRPS}}$}  &	\multicolumn{3}{c}{$\%$}	\\ 
	s	& name & SV & SC & SG &	$\text{SV} \stackrel{\%}{\succ} \text{SG}$ &	$\text{SC} \stackrel{\%}{\succ} \text{SG}$ &	$\text{SC} \stackrel{\%}{\succ} \text{SV}$	\\ 
  \hline
	55 & \texttt{alfe} & 3.20 & 2.28 & 2.59 & 0.18 & 0.78 & 0.96	\\ 
  56 & \texttt{arko} & 3.11 & 3.29 & 3.44 & 0.72 & 0.64 & 0.30	\\ 
  57 & \texttt{arns} & 2.25 & 2.47 & 2.61 & 0.79 & 0.66 & 0.25	\\ 
  58 & \texttt{augs} & 2.73 & 2.66 & 2.57 & 0.48 & 0.56 & 0.57	\\ 
  59 & \texttt{blan} & 3.00 & 2.57 & 2.64 & 0.33 & 0.63 & 0.80	\\ 
  60 & \texttt{bork} & 2.32 & 2.42 & 3.22 & 0.93 & 0.90 & 0.40	\\ 
  61 & \texttt{bvoe} & 2.32 & 2.11 & 2.59 & 0.73 & 0.84 & 0.72	\\ 
  62 & \texttt{buch} & 2.54 & 2.49 & 2.60 & 0.61 & 0.65 & 0.54	\\ 
  63 & \texttt{cosc} & 2.65 & 2.94 & 2.84 & 0.68 & 0.50 & 0.24	\\ 
  64 & \texttt{ebra} & 2.33 & 2.49 & 2.57 & 0.73 & 0.67 & 0.32	\\ 
  65 & \texttt{ellw} & 3.22 & 3.20 & 2.59 & 0.20 & 0.18 & 0.47	\\ 
  66 & \texttt{falk} & 2.64 & 2.68 & 2.61 & 0.57 & 0.56 & 0.45	\\ 
  67 & \texttt{gram} & 1.84 & 1.90 & 2.54 & 0.93 & 0.89 & 0.45	\\ 
  68 & \texttt{grue} & 1.98 & 1.92 & 2.65 & 0.91 & 0.92 & 0.60	\\ 
  69 & \texttt{luec} & 2.21 & 2.45 & 2.59 & 0.79 & 0.68 & 0.27	\\ 
  70 & \texttt{muel} & 2.22 & 2.43 & 3.03 & 0.91 & 0.86 & 0.27	\\ 
  71 & \texttt{ohrz} & 3.66 & 3.02 & 2.59 & 0.06 & 0.24 & 0.83	\\ 
  72 & \texttt{rahd} & 2.36 & 2.58 & 2.60 & 0.73 & 0.60 & 0.25	\\ 
  73 & \texttt{wies} & 2.72 & 2.81 & 2.60 & 0.45 & 0.39 & 0.38\\ 
	\hline
  mean &  & 2.59 & \textbf{2.56} & 2.71 & 0.62 & 0.64 &	0.48\\ 
  \hline
\end{tabular}
\end{table}

%\clearpage

\paragraph{Percentaged outperformance}
In addition Table \ref{tab:scores} compares the three spatial approaches by means of percentaged outperformance. For all stations of the validation data set we count the points in time for which one approach yields a better (lower) score than the other. This shows that the spatial R-vine model is preferred over the spatial Gaussian model and that the spatial local C-vine composite likelihood approach wins against the spatial Gaussian model. However there is no clear preference comparing the spatial local C-vine composite likelihood approach and the spatial R-vine model.

\paragraph{Log-score difference plots}
A log-score difference plot is a tool which helps to compare two approaches with regard to dependencies of the prediction performance on time. Figure \ref{fig:AverageScoresOP} depicts the averaged log-score (continuous ranked probability score) difference plots for the pairs 'SV-SG' and 'SC-SG'. We observe short time intervals towards the turns of the years, where the spatial Gaussian model (SG) consequently gives lower scores than the other two models. Moreover it seems, that the differences between the spatial R-vine model (SV) and the spatial Gaussian model (SG) are higher than the corresponding differences of the spatial local C-vine composite likelihood approach (SC) and the spatial Gaussian model (SG).

%% file: Sections/Conclusions.tex
\section[Conclusions and outlook]{Conclusions and outlook}\label{sec:conc}

Our model building process for the daily mean temperature data resulted in a spatial local modeling approach which is able to capture the dependencies between different locations and which allows for prediction at unobserved locations.
A comparison of the resulting spatial local C-vine composite likelihood approach ($16$ parameters) with its non-spatial complement the local C-vine composite likelihood approach ($495$ parameters) yields $1.6$ hours of computation time for the maximum composite likelihood estimation instead of $6.1$ days. Compared to the spatial R-vine model ($41$ parameters, $18$ hours of computation time) the number of parameters and the time needed for parameter estimation was also reduced. The computations were performed on a $2.6$ GHz AMD Opteron processor.

The prediction results from all three spatial approaches which are compared within this work are quite similar and reasonable, as long as we aim to predict for a location not exceeding the scope of our training data set. In order to rank the different approaches according to their prediction accuracy, we calculated and compared continuous ranked probability scores. We observed an outperformance of our two spatial, vine copula based approaches over the spatial Gaussian model.

Whereas the prediction from the spatial R-vine model uses the available information on all $54$ observation stations of the training data set, the spatial local C-vine composite likelihood approach based predictions are only conditioned on the observations of the three closest stations. These differences in the model structure obviously have an effect with regard to computation time for prediction. Therefore, further investigations of the performance of spatial local C-vine composite likelihood approaches which consider different numbers of neighboring stations should be undertaken. Moreover application of our approaches to other types of data sets are desirable. Especially data sets with asymmetric bivariate dependencies are of interest and will be studied in the future. Also multivariate extensions of our approach, e.g. joint modeling of temperature and precipitation are of interest.